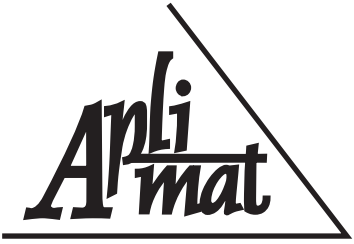

# GEOMETRY AND PERSPECTIVE IN THE LANDSCAPE OF THE SAQQARA PYRAMIDS

**MAGLI Giulio, (I)**

**Abstract.** A series of peculiar, visual alignments between the pyramids of the pharaohs of the 4th, the 5th and the 6th Egyptian dynasties exists. These alignments governed from the very beginning the planning of the funerary monuments of successive kings and, in some cases, led to establish building sites in quite *inconvenient* locations from the technical viewpoint. Explaining the topography of these monuments means therefore also investigating on their symbolic motivations: religion, power, dynastic lineage and social context, as well as getting insights on the skills of the ancient architects in astronomy and geometry. In the present paper we focus on the relationships between the Old Kingdom pyramids at Saqqara.

**Key words.** Ancient astronomy. Ancient and sacred geometry. Egyptian pyramids.

*Mathematics Subject Classification:* Primary 01A16, 51-03.

## 1 Introduction

It is known since the 19th century that an interesting feature exists in the layouts of the pyramids of the 4th dynasty at Giza: the presence of a "main axis" connecting the south-east corners of the three main monuments and directed to the area where the ancient temple of the sun of Heliopolis once stood, on the opposite bank of the Nile [1,2,3]. This line is connected with a process of "solarisation" of the pharaohs which started with Khufu, the builder of the Great Pyramid, and lasted up Menkaure, the builder of the third Giza pyramid [4,5]. The kings of the 5th dynasty moved - in spite of the presence of several favorable places to the immediate south of Giza - some 7 Kms apart on the plateau of Abusir. Here, again, a "main axis" similar to the Giza one was inaugurated: a straight line indeed connects the north-west corners of the pyramids of three successive kings [6] (on the problems related to the topography at Abusir see [7] and references therein). Besides such "diagonal" alignments, another type of visual relationship is also known to exist between pyramids and sacred sites: meridian (i.e. north-south) alignments. A meridian line was, for instance, suggested by Goyon [8] to connect Giza with the sacred center of Letopolis located due north. The research presented here is part of a wider program aimed at a complete (topographical, historical and astronomical) analysis of the whole set of alignments in the pyramid's fields of the Old Kingdom [9].

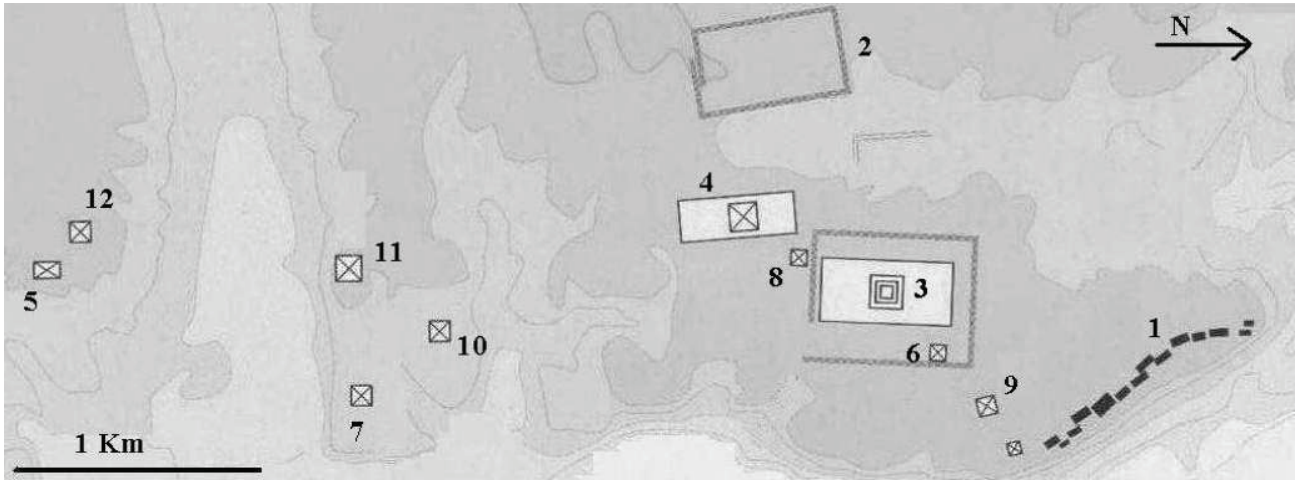

*Fig. 1 Map of Saqqara (numbering of the monuments in chronological order).*
*1- Archaic Mastabas   2-Gisr el-Mudir   3- Djoser Step Pyramid   4-Pyramid of Sekhemkhet*
*5- Mastaba El Faraun   6-Pyramid of Userkaf,   7-Pyramid of Djedkare   8- Pyramid of Unas*
*9-Pyramid of Teti   10-Pyramid of Pepi   11-Pyramid of Merenre*

The final aim of this project is to obtain a complete picture of these peculiar elements of the design of the "sacred landscape" and to analyze its implications both from the historical and from the technical point of view. In the present paper, we shall concentrate on the visual relationship existing in the pyramid's fields at Saqqara.

## 2     Historical and architectural context

In what follows we will be interested in the funerary complexes constructed during the course of the so-called Old Kingdom in the central area of Saqqara and in Saqqara south (Dashour). I will therefore recall briefly the historical development of these sites; as a reliable working framework I refer here to the chronology by Baines and Malek [10].
The "Age of the Pyramids" starts with the so-called Step Pyramid, built by the first $3^{th}$ dynasty pharaoh Djoser around 2630 BC [11]. Saqqara was already at that time a very important necropolis, located on the desert plateau overlooking the area of the capital, the "White Wall" later called Memphis (Fig. 1). Some of the pharaohs of the previous dynasties were buried in tombs located south-west of the future Djoser building site. Further, since a few years ago most Egyptologists were convinced that almost all the kings of the first two dynasties were buried in Saqqara, their tombs being the *mastabas* ("bench-like" buildings) located on a line which flanks the ridge of the plateau; today, these tombs are rather attributed to high state officials, and the tombs of the first kings are identified with the burials – before considered symbolic cenotaphs - present at Abydos, the city sacred to Osiris. This riddle is not definitively solved; in any case, the importance of Saqqara is confirmed also by the presence of another monument, pretty similar to those existing in Abydos. These monuments – usually called with the German name *Talbezirke-* are huge rectangular enclosures which were ritually leveled at the death of the king. The unique which survives intact in Abydos is the *Shunet el-Zebib*, a rectangular mud-brick building measuring 123 x 64 meters, built by king Khasekhemwy at the beginning of the 27 century BC. The (leveled) one existing in Saqqara is today called *Gisr el-Mudir*; it is roughly oriented to the points of the compass and it is located west of Sekhemkhet's pyramid. Almost certainly, it belongs to a king of the second dynasty – perhaps Khasekhemwy himself – and it was very large, with walls as wide as 15 meters and dimensions around 650x350 meters.

Djoser's pyramid complex represents a clear breakthrough in the history of art and architecture. It is comprised in a high-walled rectangular enclosure – reminiscent of the Talbezirke, but built in stone – and consists of several buildings. The key of the complex is of course the Step Pyramid, the first pyramid ever built in Egypt, about 60 meters tall. The enclosure is roughly oriented to the cardinal points (as is the pyramid) and originally had only one entrance, located near the south-east corner, and a huge surrounding dry moat [12].

After Djoser, other pyramids will be built by kings of the same dynasty. The first is Sekhemkhet's, a project very similar to Djoser's which was however left unfinished and – probably – filled with earth and rubble already by its builders. Then followed the so-called Layer Pyramid at Zawiet el Arian and another very far south, in Meidum, perhaps to be attributed to Snefru, the first king of the $4^{th}$ dynasty. In any case, it is with Snefru (around 2575 BC) that building of pyramids had the final technical breakthrough, with the first "geometrical" pyramids built with huge stone blocks and cased with Tura white limestone. These are the two magnificent pyramids located in Dashour (South Saqqara) called today the Bent Pyramid and the Red Pyramid (due to a change in the slope of the first and the color of the stones of the nucleus of the second).

With the son of Snefru, Khufu, we have the beginning of the time of the "solar kings", i.e. those who declared a direct descent from the Sun God. Khufu was the builder of the Great Pyramid in Giza (and perhaps initiated also the project of the second one [3,13,14]). His son Djedefre moved to Abu Roash, while Khafre and Menkaure built in Giza. The successor (probably the son) of Menkaure, Shepsekaf, broke however the "solar" tradition. His name does not bring the "solar" suffix -re; his funerary monument is not in view from Heliopolis and it is actually the unique royal monument of the epoch which is not a pyramid. It is indeed a somewhat unique monument, called today Mastaba El Faraun, which we shall describe in more details later (for a discussion of different viewpoints about Shepsekaf see [6]) . Perhaps inspired by a total solar eclipse occurred on April 1, 2471 in the area of the Delta, the successor Userkaf (around 2465 BC) returns to the tradition. He builds a pyramid in Saqqara, located as close as possible to the wall of the first pyramid ever constructed, Djoser's, and a huge monument devoted to the Sun God in Abu Gorab. After Userkaf, we have a new series of "solarised" kings which will successively build their pyramids in Abusir: Sahure, Neferirkare, Neferefre, Shepseskare (pyramid building site not certain) and Niuserre. After Niuserre, the kings will definitively move to Saqqara for their pyramids: Menkahour (pyramid building site not certain) and then Djedkare, who choose a prominent position directly at the ridge of the plateau. After, Unas, whose project, developed in a very unfavorable position near Djoser's, will be of particular interest for us as well as that of his follower Teti. The pyramid building site of Teti's immediate successor, Userkare, has never been individuated. The last three kings to build a pyramid before the end of the Old Kingdom will be Pepi I, Merenre and Pepi II, who all choose the area half-way between Saqqara and Dashour.

## 3    Art and landscape at Saqqara

To describe the way in which geometrical alignments and artistic perspectives were established at Saqqara we follow strictly the chronological order recalled in the previous section. Indeed, we must take into account that the whole "sacred landscape" visible today is the result of successive additions in the course of several centuries.

## 3.1 The Snefru project

The pyramids constructed by Snefru at Dashour stand, still today, among the most huge and beautiful monuments ever built by humankind. The Bent Pyramid owes its name to a sudden softening of its inclination, which was effected when the construction had reached 49 meters (the initial inclination of 54°3' drops to 43°21', thus forming an angle in the pyramid's profile). One of the ten biggest and heaviest objects ever created throughout history, the pyramid has a 189 meter wide base and is 105 meters tall. The north pyramid of Dashour or Red Pyramid owes its modern name to the reddish hue of the limestone used to build it. The base measures 218.5 by 221.5 meters, is 104.4 meters tall, with a gradient virtually identical to that of the upper section of the south pyramid. The two pyramids are relatively far apart (about 1,850 meters) and are not on the same meridian (the distance between their meridians is about 300 meters). The Red Pyramid is so far from the Nile floodplain that a two-kilometer-long causeway would have been needed to access it from the river. These facts implied several logistical difficulties (for instance, for the transport of construction materials from the Nile) and have never been satisfactorily explained. Further, at the moment of construction the whole desert area between the Saqqara central field and Dashour was almost completely free. It is therefore also unclear why Snefru choose to build his pyramids so far from the capital. To gain insight into this problem, we must first mention the fact that most Egyptologists believe that the two monuments do *not* belong to a unitary project. Generally, it is believed that the Red Pyramid was the "real" tomb and was erected because the Bent Pyramid became structurally unstable during its construction. Because the Bent Pyramid was in danger of collapsing, its gradient was first softened, and then it was decided to build a new pyramid anyway. However, not all scholars in the past agreed with this view, which is actually – at least in the author's view – simply *untenable* for a long series of reasons (a complete discussion can be found in [7]). Thus, it is much more likely to think that the Snefru project comprised two pyramids from the very beginning, as will probably occur for the project of his son Khufu at Giza [7,13,14,15]. The real explanation for the riddles of the Dashour complex are to be sought at a more symbolic level. What particularly arouses suspicion is the "duality" apparent in the site – two enormous pyramids, two (and not three) slopes, two funerary apartments in the south pyramid – prompting some scholars in the past to suggest that the pairing represents, a tribute to the tradition of the Pharaoh as the ruler of unified Upper and Lower Egypt. The "duality" in the sepulchre is actually easily traced back in the funerary cult before Snefru, e.g. in the curious "south tomb" - a underground maze similar to that beneath the Step pyramid and probably meant as a cenotaph for the king - present in Djoser's complex. The change in the slope of the Bent pyramid remains something of a mystery, however the pyramidion (the monolithic capstone which was put on the apex of the pyramids) found in pieces near the Red pyramid has the same slope of the lower part of the Bent pyramid, and was perhaps meant to introduce a further element of symmetry in the project. To this interpretation of the Snefru pyramids as a unitary project I would also add that the Snefru monuments were meant to represent an *artificial horizon.* Indeed, the ancient pathway leading from the capital to the Saqqara plateau almost certainly raised up a wadi (dried river) located a few hundreds meters to the north [11] and followed the line of the archaic mastabas up to the area were the Userkaf and Teti pyramids would later been built. At Snefru times this path was free up to the entrance of the Djoser complex on its right and a person ascending the plateau would have seen the two giant pyramids of Dashour standing alone at the profile of the horizon (actually, this impressive experience is still enjoyable today, especially in clear days). The (symbolic) horizon was, in ancient Egypt, represented in hieroglyphs by a sign of two paired mountains 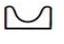. Perhaps the two *artificial* mountains of Snefru actually represented the two (re-united) parts of the country themselves. As a matter of fact, immediately thereafter, the first of the *solar* kings, Khufu, will design his funerary project at Giza following the same pattern, but adding to the „paired mountains" the sun setting in between  . This was achieved with a spectacular phenomenon, a hierophany, which can still be experienced from the Sphinx area at the summer solstice, when the sun setting between the two giant pyramids replicates one time a year in the sky the very same *name* of the Great Pyramid, *Akhet Khufu* or „the horizon of Khufu" [1,3,14].

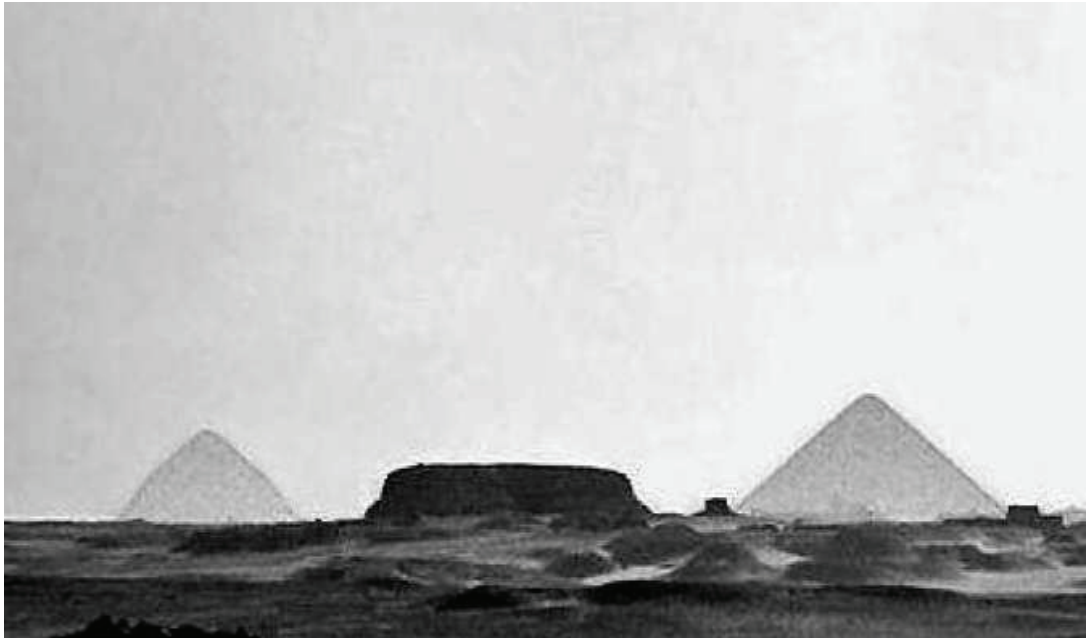

*Fig. 2 The southern horizon viewed from the the Saqqara central field, near the Teti pyramid. In the foreground, the huge mole of the Mastaba el Faraun; in the background the "Snefru project": the Bent and the Red pyramid.*

Further to such considerations, it is interesting to observe that the Snefru project was conceived taking into account the position of the pre-existing monuments of the Saqqara central field. It was indeed already noticed by Goyon [8] that the Red Pyramid lies on the same meridian of the Gisr el-Mudir. Actually, the correspondence looks closer, since the the Red and the Bent pyramid of Snefru appear to have been constructed in such a way that their west sides align with the two south corners of the enclosure. Interestingly, the Snefru project seems to be ideally connected also with Djoser's, in that the center of the valley temple of the Bent pyramid aligns with the apex of the Step Pyramid.

### 3.2 The Shepsekaf project

At the moment of construction of the funerary monument of Shepsekaf, the area between Saqqara and Dashour was still as empty as it was in Snefru's times. The king choose however to build his tomb in the very middle of this area, and therefore quite far from both the two pre-existing necropolis. Again, this fact seems to await for an explanation. To try to get rid of this enigma, we must first of all observe that the resemblance between Shepsekaf's tomb and a "true" Mastaba is only apparent. The tomb is, indeed, gigantic in size (around 100x75 meters base and 19 meters tall) and its lower courses were cased in granite, as were those of the pyramid of his father Menkaure; further, the interior arrangement of it is that of a pyramid. Contrary to mastabas indeed – which were usually composed by a sort of "apartment" of several rooms devoted to the funerary rites, while the funerary chamber was located in deep shaft, the Mastaba el Faraun has a descending corridor oriented to true north and lined in granite, and a subterranean antechamber/chamber system. Also the shape of the "bench" is much more elaborate to that of a Mastaba, with a sort of vault between vertical ends; it has been proposed that it recalls a giant sarcophagus or archaic models such as the so-called Buto shrine [11]. Actually, at least in the author's view, the monument also resembles the hieroglyph for "sky" ⌐⌐. All in all, we can be certain that also the Mastaba el Faraun was conceived – from the point of view of the art of the landscape – as a monument *devoted to occupy the horizon*. Of course, the south horizon from Saqqara was already "occupied" by the Snefru pyramids as Verner [6] has already put in evidence, suggesting that Shepsekaf's choices might have been motivated by the will of the king to exploit his dynastic lineage up to Snefru. My proposal is that it was indeed the complete project – choice of the building site, and design – to be conceived in order of harmonize the monument with the pre-existing Snefru-built landscape, and therefore stressing the return of the king to a "pre-solar", "horizon" tradition. Indeed, if a line is traced from half the distance between the two Snefru pyramids and the center of Shepsekaf, it is seen that it crosses the Saqqara central field in the same "entrance" area located near the Teti pyramid. As a consequence, anyone reaching the summit of the ridge would have seen (and still can see, Fig. 2) the king's tomb forming a sort of regular baseline for the double-mountains symbol "created" at the horizon by the two giant pyramids of Snefru; on the other end, it is easily seen that the position of the monument is not dictated by the morphology of the territory: the huge building is founded on an artificial terrace and is relatively far from the ridge of the Plateau.

### 3.3 The Unas project

As mentioned in the previous section, the pyramid building sites of the solarized kings (Abu Roash, Giza and Abusir) exhibit the "dynastic lineage" of the kings from the Sun God by means of topographical alignments with Heliopolis [2]. In particular, the south-east corners of the three pyramids in Giza and the north-west corners of three pyramids in Abusir align in the direction of the sacred city. Interestingly enough, the idea of "dynastic" alignments (representing lineage, or closeness of religious ideas, or more simply hints to past traditions) remained up to the end of the Pyramids age and is clearly visible also at Saqqara. There is, in fact, no other feasible way to explain the most "crazy" pyramid complex ever built, that of king Unas, the last king of the 5[th] dynasty. The pyramid is constructed near the south-west corner of the precinct of Djoser, thus very far in the desert. Consequently, the builders had to construct also a very long (more than 700 meters) causeway connecting the complex with the valley temple. Even worse, they had to "clear" the zone near Djoser's south wall which was already overcrowded by many pre-existing mastabas and even by some royal tombs of the first dynasties. Some tombs were thus filled with earth, and some mastabas were even completely dismantled (one has been reconstructed in the 70' of last century from blocks found beneath the causeway). Clearly, Unas must have had an important reason for choosing such an unfavorable building site, if only one considers that the pyramid could have been built *in front* of the area south of the Step pyramid (admitting that the king wanted to stay close to it anyway). Actually, already in 1985 Lehner [1] noticed the existence of a "Saqqara diagonal", without however attempting to discuss its meaning. It is a line oriented roughly SW-NE (it is difficult to ascertain the azimuth precisely, but it can be estimated as being ~39° east of north) which connects the south-east corner of Userkaf pyramid with the south-east corner of Djoser's pyramid and then crosses over Unas' north-east/south-west base diagonal. There can be no possible doubt, at least in the present author's view, that it was precisely the will of realizing this alignment which governed the choices made by Unas' architects. Actually indeed, the Unas pyramid has also another "strange" peculiarity. Although its side base (57.7 m) is the smallest among all the royal pyramids – implying a relevant economy in its construction - the monument is also the one with the steepest slope (the slope was of course chosen with the help of rational fractions, and Unas' is 3/2 [16]). This implies that the pyramid rose at the respectable height of 43 meters. The likely reason for this choice is that another problem in realizing the project was that it was to be recognizable to anyone approaching the plateau, while the height of the wall of the Step pyramid complex would have obstructed the view of the pyramid (on the issue of visibility see also the discussion). To increase visibility further, the alignment was designed along the diagonal of the pyramid, not along the south-east corner, again a unique case among the various "diagonals". The final result of the Unas project is that the placement of the three pyramids of the Saqqara central field resembles that of the three pyramids of Giza, to the point that the resemblance can hardly be considered casual. A similarity was already noticed some years ago by Goedicke [17], who observed that an unobstructed line of sight connects the Userkaf pyramid with Khufu's.

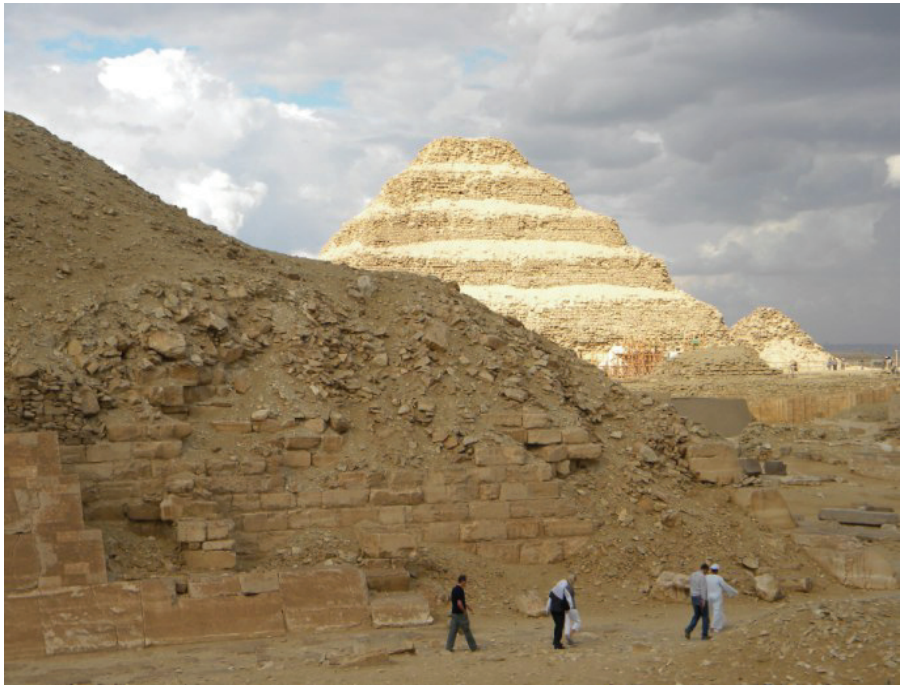

*Fig. 3 A picture from south of Unas pyramid (foreground). The "Saqqara diagonal" connecting Unas' diagonal with the south-east corners of the Step (middle) and Userkaf (background) pyramids can be perceived .*

Actually, if connecting lines are traced between the summits of Djoser's and Unas' pyramids and the apexes of Khafre's and Menkaure's monuments respectively, it becomes clear that the placement of Unas pyramid was actually conceived to realize a sort of (rough) copy of the arrangement of the Giza pyramid field. These lines are in fact about 14.5 Kms long (and therefore allow a direct visibility) and, although they are not parallel, their relative deviation stays within 2°.

### 3.4 The Teti project

Lehner [1] noticed that the Saqqara diagonal also touches (roughly) the north-west corner of Sekhemkhet's unfinished pyramid to the south-west, and Teti's north-west corner to the north-west. Perhaps, however, the alignment with Sekhemkhet's pyramid occurs by chance, since the pyramid itself might have been already buried at Unas' time [9]. Teti instead was Unas' successor, and it is therefore conceivable that he may have wanted to align his pyramid along the "Saqqara diagonal".

The architect of the king was faced, however, with a difficult problem, since it was nearly impossible to build the pyramid in a position very far in the desert, along the diagonal to the south-west of the Unas one, while the outcrop of the Saqqara ridge in front of Userkaf pyramid was completely free. So, Teti complex was located in this area. Interestingly enough, in the course of the project it was decided to try to respect the "dynastic" perspective of the pre-existing diagonal, and probably for this reason it is the north-west corner of the Teti pyramid to lie on the pre-existing line. Another "mystery" of the Teti project is the fact that the pyramid complex is badly aligned with respect to the cardinal points (it deviates about 9 degrees west of north). Recently however, it has been shown that is probably due to a solar (as opposed to stellar) orientation of the complex [18].

### 3.5 The pyramids of the 6th dynasty

The pyramid building site of Teti's immediate successor, Userkare, has never been individuated. His reign was very short, and perhaps his project remained unfinished; in what follows we shall, however, make a guess on the place where its remains could perhaps been found. The successive kings of the 6th dynasty, Pepi I, Merenre and Pepi II, choose indeed the area between the Saqqara main field and Shepsekaf's monument, where only Djedkare monument was pre-existing.

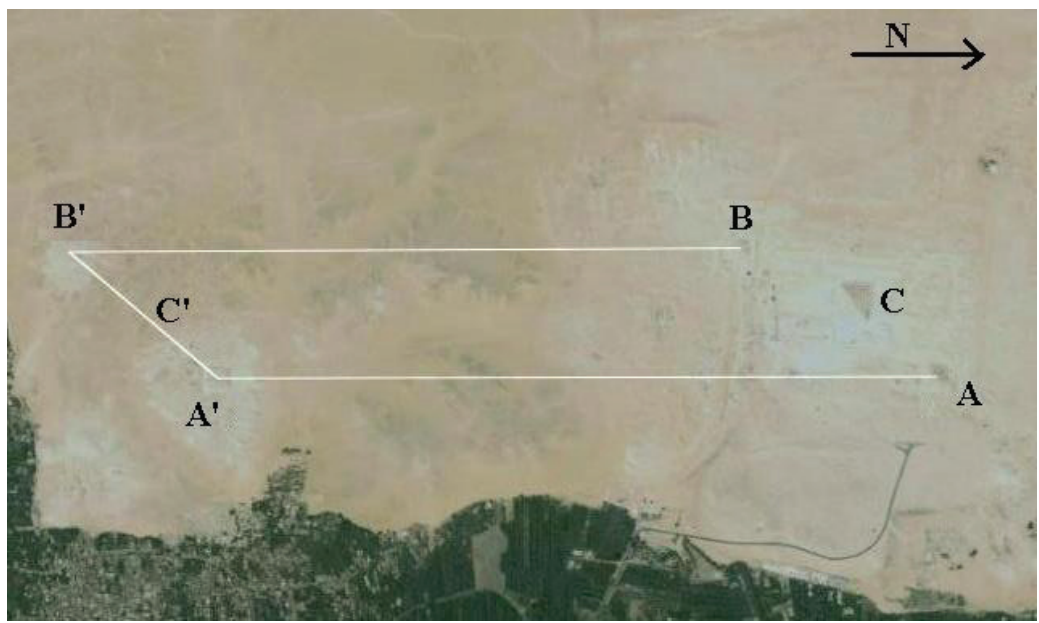

*Fig. 4 Meridian alignments between the Saqqara central group and the 6th dynasty pyramids. A- Userkaf A'- Pepi I; B- Unas B' -Merenre; C denotes the Step Pyramid, C' the hypothetical position of a further royal project, perhaps Userkare's. The diagonal Pepi I-Merenre is also shown.*

In particular, Pepi I built his pyramid near the ridge of the plateau. A meridian alignment connects the apex of his pyramid with that of Userkaf in the central Saqqara field. This may be considered as a chance. However, if we look at the position of his successor Merenre we see that he moved south-west in such a way to align the diagonal to that of Pepi I and the center of the pyramid with the apex of Unas (this line also touches the west side of Shepsekaf's Mastaba to the south). Therefore, there is a symmetrical connection between the Pepi I/Merenre complexes and the complexes placed at the two ends of the Unas diagonal, namely Unas and Userkaf. This connection can hardly be considered as casual, since it seems – once again – that Pepi I and Merenre wanted to replicate a pre-existing "diagonal", in this case the Saqqara diagonal. If this is accepted, then it is immediately seen that the position at the center of the diagonal, which corresponds to a meridian alignment with the Step pyramid, is left free. One may at this point seriously suspect that the – perhaps unfinished – complex of Userkare may have been the first to be located in this area *and might therefore be located approximately in the middle of the line connecting Pepi I and Merenre's diagonals*. Interestingly, recent excavations in the nearby Tabbet al-Guesh area gave many hints at the possibility that a pyramid has still to be discovered there [19]. The area is actually criss-crossed by yet another possibly non-casual connecting line, since the northwest-southeast diagonal of Pepi I aligns with the south-east corner of the Gisr el-Mudir. Regarding Pepi II, successor of Merenre, we shall never know if he would have liked to add his monument to the south-west of Merenre's because that area is occupied by a dried river and is unsuitable for building. Perhaps as a consequence, he choose a position immediately to the south, near Shepsekhaf's monument.

## 4       Discussion and conclusion

The placement of pyramidal complexes had to take into account a series of practical factors such as presence of nearby stone outcrops to be transformed into quarries and accessibility of materials. Perhaps also the presence of the - still active - building site of the pyramid of the preceding pharaoh may have influenced the choice [20]. However, there can be no doubt on the fact that, in many cases, the pyramids were *not* constructed were reasonableness would have wanted. Following the historian of religion Mircea Eliade, according to whom [21] the symbolism contained in the sacred space is so "old and familiar" that it may be difficult to recognize it, we have thus pursued a systematic search of the possible geometrical and perceptive connections between the pyramids of the Old Kingdom. What turns out is, that it was the way in which the pyramidal complex harmonized with the pre-existing landscape to be the *main* motivation for many of the topographical choices. Here, „landscape" must be understood in a very broad sense: it was the natural landscape, the sky, and also what could be called „dynastic" landscape, namely the will of putting the funerary monument in direct relationship with pre-existing ones, built by pharaohs which were related by direct lineage (e.g. as occurs in Giza and Abusir) and/or by closeness of religious/political ideas. From this point of view, the results of the present research have still to be developed, and may lead to historical information about poorly known pharaohs such as Shepsekaf.
A point which emerges clearly is, that it was not enough that the planners or perhaps the priests were aware of the existence of a symbolic link of the newly built pyramid complex with the pre-existing ones. These links were by no means a sort of „esoteric" (or worse „initiatic") knowledge: they had to be made visible - I would say *familiar* - to any pious person approaching the royal Necropolis, the places were the cults of the dead pharaohs was carried out. Actually, the ideal lines which visually connect such sacred places are still clearly perceptible, after 4500 years, to anyone visiting the pyramids fields today.
The existence of these lines raises also a series of technical issues about ancient survey techniques and astronomy. Indeed, from one side, alignments with pre-existing, elevated points put in evidence with sun-reflecting (golden) signals might have been useful to establish references during the planning and the construction of the pyramids; therefore, alignments - especially meridian - may have had *also* a technical motivation. In any case, how were they obtained? To answer this question it is necessary to assess the precision attained in them. The validity of the meridian alignments mentioned here has been controlled on topographic maps and, whenever possible, by direct survey with a precision magnetic compass, with a nominal precision of 0.5°. However, when the same alignments are re-run using the Google-earth program, they are not only fully confirmed, but most of them turn out to be valid with an astonishing precision, comparable to that reached by the ancient

Egyptians in orienting the pyramids (of the order of 20' or even less). This would lead to new possibilities in checking ideas about orientation of monuments on the Old Kingdom, like e.g. the so-called simultaneous transit theory (a complete discussion of these theories can be found in [7]). The satellite images used by the program can produce distorsional effects however, and an error is always to be expected in centering the monuments. It would, therefore, be very interesting to carry out complete survey of these alignments using a high-precision transit instrument, to check if this computer-based estimate is truly valid.

Finally, an interesting issue is connected with the ubiquitous (Giza, Niuserre at Abu Sir, Unas at Saqqara, Saqqara south) existence of survey lines oriented (more or less precisely) quarter-cardinally. A similar family of orientations has been shown to exist for the Egyptian temples [14]. These orientations are difficult to explain, and it has been proposed that it was achieved by determining the meridian and turning its direction by 45º; the origin of such a custom should be the will of accomplishing two religious precepts: orientation to the celestial realm (the circumpolar stars) and orientation perpendicular to the Nile [18]. However, the idea of "averaging" between these two conflicting needs looks unconvincing. For instance, it certainly cannot be applied to Mesopotamia, where similar customs of „quarter cardinal" orientations entered in use already in the 4$^{th}$ millennium BC (e.g. at Eridu) [22]. Further, it can hardly be applied to alignments of pyramids, since all such monuments took already care of both needs in having entrances aligned to true north *and* valley temples facing the Nile. Actually, the first of inter-cardinal diagonals, that connecting Giza with Heliopolis, corresponds with impressive precision to the setting of the brightest part of the Milky Way. At the times of the construction of the Great Pyramid (say 2550 BC) an observer looking from Heliopolis would have seen the stars of the Southern Cross-Centaurus "flow" together with the celestial river – probably the „Winding Waterway" cited in the Pyramid Texts - and disappear from view behind the apex of Khufu's monument. Due to the extension of our galaxy as a sky-band, this interpretation (with the corresponding one at rising) remains - in spite of precession effects – a valid alternative for inter-cardinal orientations during the Old Kingdom; due to the similar latitudes, the same explanation might be hypothesized for Mesopotamian temples as well.

**Current address**

**Giulio Magli,**
Dipartimento di Matematica del Politecnico di Milano, P.le Leonardo da Vinci 32,
20133 Milano, Italy.
e-mail Giulio.Magli@polimi.it